\documentclass[12pt,fleqn]{article}
\usepackage{graphicx,amssymb,epsfig,times}

\def\be{\begin{equation}}
\def\ee{\end{equation}}
\def\bea{\begin{eqnarray}}
\def\eea{\end{eqnarray}}
\def\bean{\begin{eqnarray*}}
\def\eean{\end{eqnarray*}}
\def\bary{\begin{array}}
\def\eary{\end{array}}
\def\nn{\nonumber}

\def\lbar{\overline}
\def\lsim{\mathrel{\rlap{\lower3pt\hbox{$\sim$}}\raise2pt\hbox{$<$}}}
\def\gsim{\mathrel{\rlap{\lower3pt\hbox{$\sim$}}\raise2pt\hbox{$>$}}}

\begin{document}

\begin{titlepage}

\begin{flushright}
  hep-ph/0502183 \\
  February 2005
\end{flushright}
\bigskip
\bigskip

\begin{center}
  
  {\boldmath\bf\Large Constraining $\gamma$ from $K^* \pi$ and $\rho K$ Decays
    \footnote{To be submitted to Phys.~Rev.~D.}
  } \\
  \bigskip \bigskip
  {\sc Cheng-Wei Chiang \footnote{E-mail address: chengwei@phy.ncu.edu.tw}} \\
  \bigskip
  {\it Department of Physics, National Central University, Chungli,
    Taiwan 320, R.O.C. \\
    and \\
    Institute of Physics, Academia Sinica, Taipei, Taiwan 115, R.O.C.}

\end{center}

\date{\today}

\begin{abstract}
  We show that information on the weak phase $\gamma$ can be extracted from the
  $K^* \pi$ and $\rho K$ decays.  Less hadronic uncertainty is involved when
  the observables of four of these modes are combined together.  We further
  point out two approximate relations in these decay modes which can help
  determine whether there are new physics contributions in $\Delta I = 1$
  transitions, as hinted in the $K \pi$ modes.
\end{abstract}

\leftline{\qquad PACS codes: 12.15.Hh, 13.25.Hw, 11.30.Er}

\end{titlepage}


\section{Introduction}
\label{sec:intro}

$B$ meson decays have been a rich source of useful information on the
Cabibbo-Kobayashi-Maskawa (CKM) mechanism in the standard model (SM).  It is
now an active field to take advantage of recent and future data from
$B$-factories to accurately fix the shape of the unitarity triangle of the CKM
matrix.  In particular, we have been able to determine in recent years one of
the angles, $\beta$ (or $\phi_1$), to a high accuracy using mainly $b \to c
\bar{c} s$ decays \cite{Abe:2004mz,Aubert:2004zt}.  It is therefore of great
interest to find reliable methods for determining the other angles as well.
Several methods have been proposed to determine the angle $\gamma$ (or
$\phi_3$) using hadronic $B$ decays, such as the $DK$ modes
\cite{Gronau:1990ra}, the $K \pi$ modes \cite{Neubert:1998pt,Gronau:2001cj},
the $\eta \pi$ modes \cite{Chiang:2003rb}, the $K^{*\pm} \pi^{\mp}$ modes
\cite{Sun:2003wn}, and the $\pi\pi$ and $K \pi$ modes \cite{Charng:2005bz}, by
combining the data of branching ratios and $CP$ asymmetries.  However, recent
data suggest possible new physics contributions to the $K \pi$ decays and
therefore cast some doubt on the relibility of $\gamma$ thus obtained
\cite{Gronau:2003kj,Buras:2003dj,Barger:2004hn,Khalil:2004yb,Mishima:2004um,%
Baek:2004rp}.

With accumulating data it is now possible to consider an alternative method,
which employs the $K^* \pi$ and $\rho K$ decays to constrain $\gamma$.  These
$VP$ decay modes are closely related to their $PP$ counterparts, the $K \pi$
modes, where $P$ and $V$ denote respectively pseudoscalar and vector mesons,
because of their similar flavor structures.  However, they do differ in that
the final state mesons in the $VP$ decays contain different polarization
components, whereas the $PP$ decays do not.  Moreover, the $VP$ decay
amplitudes can be divided into two types: those in which the spectator quark
goes to the pseudoscalar meson in the final state and those in which the
spectator quark goes to the vector meson.

Consequently, the $VP$ decays involve two disjoint sets of contributing
topological amplitudes.  One connection between these sets of amplitudes is by
Lipkin's assumption which states that the penguin amplitude in the $K^* \pi$
modes, $P'_P$, is equal to that in the $\rho K$ modes, $P'_V$, in magnitude but
opposite in sign based on a parity argument \cite{HJLP}.  This assumption can
be used to readily explain the observed $K^* \eta$, $K^* \eta'$ branching
ratios as a result of interference between these two types of penguin
amplitudes.  It is also verified in global $\chi^2$ fits to observed decay data
in the framework of flavor SU(3) symmetry \cite{Chiang:2003pm}.

This paper is organized as follows.  We provide the necessary formulas and
current data for $B \to K^* \pi$ and $\rho K$ modes in Section
\ref{sec:basics}.  We first consider the constraints on $\gamma$ from
individual sets of $K^* \pi$ and $\rho K$ decays in Sections \ref{sec:Kstarpi}
and \ref{sec:rhoK}, respectively.  In Section \ref{sec:eqset} we show that
information on $\gamma$ with less hadronic uncertainty can be obtained by
combining the four $K^* \pi$ and $\rho K$ observables when Lipkin's assumption
is used.  We discuss possible improvements in Section \ref{sec:others} and
conclude our findings in Section \ref{sec:conclusions}.

\section{Basics}
\label{sec:basics}

\begin{table}[t]
\caption{Experimental data measured by BaBar, Belle, and CLEO collaborations
  for $B \to K^* \pi$ and $\rho K$ modes.  The branching ratios are quoted
  in units of $10^{-6}$.  The $CP$ asymmetry of a decay mode, if measured, is
  quoted in the second line.
\label{tab:dS1data}}
\begin{center}
\begin{tabular}{lllll}
\hline\hline
 & Mode & BaBar & Belle & CLEO \\
\hline
$B^+ \to$
    & $K^{*0} \pi^+$
        & $10.5 \pm 2.0 \pm 1.4$ \cite{Aubert:2004fn}
        & $9.83 \pm 0.90 ^{+1.06}_{-1.24}$ \cite{Garmash:2004wa}
        & $7.6^{+3.5}_{-3.0}\pm1.6 \; (<16)$ \cite{Jessop:2000bv}
        \\
    & $K^{*+} \pi^0$
        & -
        & -
        & $7.1^{+11.4}_{-7.1}\pm1.0 \; (<31)$ \cite{Jessop:2000bv}
        \\
    & $\rho^0 K^+$
        & $5.2 \pm 1.2 \pm 0.7$ \cite{Aubert:2004fn}
        & $4.78 \pm 0.75 ^{+1.01}_{-0.97}$ \cite{Garmash:2004wa}
        & $8.4^{+4.0}_{-3.4}\pm1.8 \; (<17)$  \cite{Jessop:2000bv}
        \\
    & $\rho^+ K^0$
        & -
        & -
        & $<48$ \cite{Asner:1996hc}
        \\
\hline
$B^0 \to$
    & $K^{*+} \pi^-$
        & $11.9 \pm 1.7 \pm 1.1$ \cite{Aubert:2004uf}
        & $14.8^{+4.6+2.8}_{-4.4-1.3}$ \cite{Chang:2004um}
        & $16^{+6}_{-5}\pm2$ \cite{Eckhart:mb}
        \\
        &
        & $-0.04\pm0.13$ \cite{Aubert:2004uf}
        & -
        & $0.26^{+0.33+0.10}_{-0.34-0.08}$ \cite{Eisenstein:2003yy}
        \\
    & $K^{*0} \pi^0$
        & $3.0 \pm 0.9 \pm 0.5$ \cite{Aubert:2004bt}
        & $0.4^{+1.9}_{-1.7} \pm 0.1$ \cite{Chang:2004um}
        & $0.0^{+1.3+0.5}_{-0.0-0.0} \; (<3.6)$ \cite{Jessop:2000bv}
        \\
        &
        & $-0.01^{+0.24}_{-0.22} \pm 0.13$ \cite{Aubert:2004bt}
        & -
        & -
        \\
    & $\rho^- K^+$
        & $8.6 \pm 1.4 \pm 1.0$ \cite{Aubert:2004bt}
        & $15.1^{+3.4+2.4}_{-3.3-2.6}$ \cite{Chang:2004um}
        & $16.0^{+7.6}_{-6.4} \pm 2.8 \; (<32)$ \cite{Jessop:2000bv}
        \\
        &
        & $0.13^{+0.14}_{-0.17} \pm 0.14$ \cite{Aubert:2004bt}
        & $0.22^{+0.22+0.06}_{-0.23-0.02}$ \cite{Chang:2004um}
        & -
        \\
    & $\rho^0 K^0$
        & $5.1 \pm 1.0 \pm 1.2$ \cite{Aubert:2004kk}
        & $<12.4$ \cite{Huang:2002ev}
        & $<39$ \cite{Asner:1996hc}
        \\
\hline\hline
\end{tabular}
\end{center}
\end{table}

\begin{table}[t]
\caption{Topological decompositions and averaged experimental data for $B \to
 K^* \pi$ and $\rho K$ modes.  Scale factors are given in the parentheses.
  Amplitude magnitudes $|A_{\rm exp}|$ extracted from experiments are quoted in
  units of eV.  Exchange, annihilation, and color-suppressed EW penguin
  diagrams are neglected by assuming their dynamical suppression.
\label{tab:dS1}}
\begin{center}
\begin{tabular}{llcccc}
\hline\hline
 & Mode & Amplitudes 
 & BR ($\times 10^{-6}$)
 & $|A_{\rm exp}|$ & $A_{CP}$ \\ 
\hline
$B^+ \to$
    & $K^{*0} \pi^+$
        & $P'_P$
        & $9.76 \pm 1.19$  & $32.6 \pm 2.0$ & - \\
    & $K^{*+} \pi^0$ 
        & $-\frac{1}{\sqrt{2}}(P'_P + P'_{EW,V} + T'_P + C'_V)$
        & $<31$  & $<58.1$ & - \\
    & $\rho^0 K^+$
        & $-\frac{1}{\sqrt{2}}(P'_V + P'_{EW,P} + T'_V + C'_P)$
        & $5.15 \pm 0.90$ & $23.7\pm2.1$ & - \\
    & $\rho^+ K^0$
        & $P'_V$
        & $<48$ & $<72.3$ & - \\
\hline
$B^0 \to$
    & $K^{*+} \pi^-$
        & $-(P'_P + T'_P)$ 
        & $12.7 \pm 1.8$ & $38.6 \pm 2.7$ & $-0.00 \pm 0.12$ \\
    & $K^{*0} \pi^0$
        & $\frac{1}{\sqrt{2}}(P'_P - P'_{EW,V} - C'_V)$
        & $1.69 \pm 1.01~(S=1.34)$ & $14.1 \pm 4.2$ & $-0.01 \pm 0.26$ \\
    & $\rho^- K^+$
        & $-(P'_V + T'_V)$
        & $9.85 \pm 1.85~(S=1.19)$ & $34.0 \pm 3.2$ & $0.17 \pm 0.15$ \\
    & $\rho^0 K^0$
        & $\frac{1}{\sqrt{2}}(P'_V - P'_{EW,P} - C'_P)$
        & $5.1 \pm 1.6$ & $24.5 \pm 3.7$ & - \\
\hline\hline
\end{tabular}
\end{center}
\end{table}

We collect the latest $CP$-averaged branching ratio and $CP$ asymmetry data of
the relevant decays in Table~\ref{tab:dS1data}.  The decay properties can be
studied in the topological amplitude formalism
\cite{Zeppenfeld:1980ex,Savage:1989ub,Chau:1990ay,Gronau:1994rj}.  In
Table~\ref{tab:dS1}, we list the topological amplitude decompositions of the
$K^* \pi$ and $\rho K$ decays \cite{Dighe:1997wj} along with the averaged decay
strengths and $CP$ asymmetries compiled from Table~\ref{tab:dS1data}.  When
computing each invariant decay amplitude from the corresponding branching
ratio, we have first used the central values of the $B$ meson lifetimes:
$\tau(B^+) = (1.653 \pm 0.014)$ ps and $\tau(B^0) = (1.534 \pm 0.013)$ ps
\cite{HFAG}, and the two-body $B$ decay formula
\be
\label{eq:width}
\Gamma(B \to M_1 M_2)
= \frac{p_c}{8 \pi m_B^2} |{\cal A}(B \to M_1 M_2)|^2 ~,
\ee
where $p_c$ is the magnitude of the 3-momentum of the final state meson in the
rest frame of $B$, $m_B$ is the $B$ meson mass, and $M_1$ and $M_2$ can be
either pseudoscalar or vector mesons.

In view of the much suppressed exchange, annihilation, and color-suppressed EW
penguin amplitudes, we only keep the tree ($T'$), color-suppressed tree ($C'$),
penguin ($P'$), and color-allowed electroweak (EW) penguin ($P'_{EW}$)
amplitudes in the table.  In our notation, the primes refer to the amplitudes
of $|\Delta S| = 1$ decays while the unprimed ones are reserved for $\Delta S =
0$ decays.  The subscript $P$ ($V$) associated with the amplitudes denotes that
in the process the spectator quark in the $B$ meson ends up in the pseudoscalar
(vector) particle in the final state.  All these processes are dominated by
penguin amplitudes.

Since the branching ratio of the $\rho^+ K^0$ mode has not been measured, we
will take Lipkin's assumption $P'_V = -P'_P$ to continue the analysis whenever
necessary.  However, such an assumption can be relaxed or corrected accordingly
once the decay is observed.

In light of their simplicity, we first consider the following ratios
\bea
R(K^* \pi)
&\equiv& \frac{\lbar\Gamma(K^{*+} \pi^-)}{\lbar\Gamma(K^{*0} \pi^+)}
= \left| \frac{P'_P + T'_P}{P'_P} \right|^2
= 1.40 \pm 0.26 ~, \\
R(\rho K)
&\equiv& \frac{\lbar\Gamma(\rho^- K^{+})}{\lbar\Gamma(\rho^+ K^{0})}
= \left| \frac{P'_V + T'_V}{P'_V} \right|^2
> 0.22 \nn \\
\label{eq:RrhoKdef}
&\to& \frac{\lbar\Gamma(\rho^- K^{+})}{\lbar\Gamma(K^{*0} \pi^+)}
= \left| \frac{P'_V + T'_V}{P'_P} \right|^2
= 1.09 \pm 0.24 ~,
\eea
where $\lbar \Gamma$ refers to the $CP$-averaged decay width and ``$\to$'' in
the second line of Eq.~(\ref{eq:RrhoKdef}) indicates where Lipkin's assumption
is used.  The numerical values are computed from Table~\ref{tab:dS1}.  These
ratios can be expressed, as will be seen later, in terms of relative strong and
weak phases, along with the parameters
\begin{eqnarray}
\label{eq:r1def}
r_1 &\equiv& \left| \frac{T'_P}{P'_P} \right| ~, \\
\label{eq:r2def}
r_2 &\equiv& \left| \frac{T'_V}{P'_V} \right| ~.
\end{eqnarray}

Since factorization works well in the tree amplitudes, $T'_{P,V}$ can be
related to $T'$ in the $|\Delta S| = 1$ two-pseudoscalar $B$ meson decays:
\bea
r_P
&\equiv& \frac{T'_P}{T'}
= \frac{f_{K^*} F_1^{B\pi}(m_{K^*}^2)}{f_K F_0^{B\pi}(m_K^2)}
\simeq 1.43 ~, \\
r_V
&\equiv& \frac{T'_V}{T'}
= \frac{A_0^{B\rho}(m_K^2)}{F_0^{B\pi}(m_K^2)}
\simeq 0.85 ~,
\eea
where the numerical results are evaluated using the BSWII form factor model
\cite{BSWII}, $f_\pi = 130.7$ MeV, $f_K = 159.8$ MeV, and $f_{K^*} = 217$ MeV
\cite{Hagiwara:fs}.  Moreover, assuming flavor SU(3) symmetry, $T'$ can be
related to $T$ extracted from the $\Delta S = 0$ decays by simple CKM factor
and decay constant ratio:
\be
\frac{T'}{T}
\simeq \frac{\lambda}{1 - \lambda^2 /2} \frac{f_K}{f_{\pi}}
= 0.284 ~,
\ee
where we have used $\lambda = 0.2265$ \cite{HFAG}.  A recent analysis
\cite{Luo:2003hn} of the semileptonic $B \to \pi \ell \nu$ decay yields $|T| =
24.4^{+3.9}_{-1.2}$ eV.  In the current analysis, we take $|T| = 24.4 \pm 7.6$
(the error taken to be 1.96 times the upper error of the previous number) eV as
a conservative estimate to take into account model dependence in the extraction
of Ref.~\cite{Luo:2003hn}, later scaling, and the part of penguin contributions
with the same CKM factor as the tree amplitude (although this part is not
significant for explaining the data, as shown in Ref.~\cite{Chiang:2003pm}).
We thus obtain $|T'_P| = 10.0 \pm 3.1$ eV, $|T'_V| = 5.9 \pm 1.9$ and
\begin{eqnarray}
\label{eq:r1}
r_1 &\simeq& 0.32 \pm 0.10 ~, \\
\label{eq:r2}
r_2 &\simeq& 0.19 \pm 0.06 ~,
\end{eqnarray}
where $|P'_P|$ is directly extracted from the branching ratio of $B \to K^{*0}
\pi^+$ and $|P'_V|$ is temporarily using the same value.  For comparison, $r_1
= 0.37 \pm 0.03$ and $r_2 = 0.26 \pm 0.03$ as determined from a global fit to
the $VP$ modes \cite{Chiang:2003pm}.

\section{The $K^* \pi$ Modes}
\label{sec:Kstarpi}

Let's concentrate on the $K^* \pi$ decay modes in this section.  Using the
parameter $r_1$, we have the decay amplitude
\be
{\cal A}(K^{*+} \pi^-)
= - P'_P \left[ 1 - r_1 e^{i (\delta_P + \gamma)} \right] ~, \\
\ee
where we fix the phase convention that $P'_P = - |P'_P|$ and $\delta_P$ is the
strong phase of $T'_P$ relative to the real axis.  It is noted that a small
penguin amplitude with the same weak phase as the tree amplitude is absorbed
into it.  This then gives
\begin{eqnarray}
\label{eq:RKstarpi}
R(K^{*+} \pi^-)
&=& 1 - 2 r_1 \cos\delta_P \cos\gamma + r_1^2 ~, \\
\label{eq:RACPKstarpi}
R(K^{*+} \pi^-) A_{CP}(K^{*+} \pi^-)
&=& - 2 r_1 \sin\delta_P \sin\gamma ~.
\end{eqnarray}
Eliminating the $\delta_P$ dependence in Eqs.~(\ref{eq:RKstarpi}) and
(\ref{eq:RACPKstarpi}) gives
\begin{equation}
\label{eq:contour}
\left( \frac{1 + r_1^2 - R(K^* \pi)}{2 r_1 \cos\gamma} \right)^2
+ \left( \frac{R(K^* \pi) A_{CP}(K^* \pi)}{2 r_1 \sin\gamma} \right)^2
= 1 ~.
\end{equation}
Given fixed values of $r_1$ and $A_{CP}(K^* \pi)$, one can obtain a curve that
relates $\gamma$ to $R(K^* \pi)$.  It is noted that the above equations are
invariant under the transformations $(\gamma,\delta_P) \to (\pi \pm \gamma,\pi
\pm \delta_P$).  Thus, there is a four-fold ambiguity in the extraction of
$\gamma$.  Here we will restrict ourselves to the solution only in the first
quadrant in view of the $39^\circ - 80^\circ$ range at $95\%$ confidence level
extracted from other observables \cite{Hocker:2001xe}.  Since only the absolute
value of $A_{CP}(K^* \pi)$ matters in Eq.~(\ref{eq:contour}), the allowed
$\gamma$-$R(K^* \pi)$ region shall be bounded in part by the curves
corresponding to $|A_{CP}(K^* \pi)|_{\rm min} = 0$ and $|A_{CP}(K^* \pi)|_{\rm
  max} = 0.13$.

\begin{figure}[t]
\centerline{\includegraphics[width=8cm]{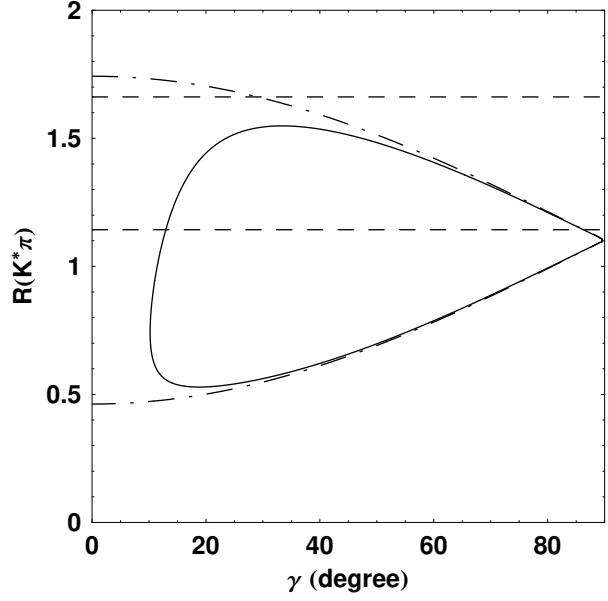}}
\caption[]{Relation between measured $R(K^* \pi)$ and the weak phase $\gamma$
  for $r_1 = 0.32$.  The solid curve corresponds to $|A_{CP}(K^* \pi)| = 0.13$,
  and the dash-dotted curve corresponds to $A_{CP}(K^* \pi) = 0$.  The dashed
  lines represent the $1\sigma$ range of the observed $R(K^* \pi)$.
\label{fig:RKstarPi}}
\end{figure}

For illustration purposes, we fix $r_1 = 0.32$ in Fig.~\ref{fig:RKstarPi}.  The
solid and dash-dotted curves correspond respectively to $|A_{CP}(K^* \pi)| =
0.13$ and $A_{CP}(K^* \pi) = 0$.  An upper bound of $\gamma \lesssim 86^\circ$
is seen in the drawing.  The lower bound is determined by the $|A_{CP}(K^*
\pi)|_{\rm min}$ curve and shown to be trivial according to current data.  The
intersection vertex of the these curves at $\gamma = 90^\circ$ rests at the
lower end of or below the $1\sigma$ range of $R(K^* \pi)$ for all possible
$r_1$ within its $1\sigma$ limits of Eq.~(\ref{eq:r1}).  Therefore, more
conservative upper bounds on $\gamma$ are obtained for larger values of $r_1$.
However, certain values of $\gamma$ in the middle range may be disfavored if
the $|A_{CP}(K^* \pi)|_{\rm max} = 0.13$ curve exceeds the upper boundary of
$R(K^* \pi)$.  For example, the range $18^\circ - 53^\circ$ is disfavored when
$r_1 = 0.42$.  If one uses $r_1 = 0.23$ as determined later in Section
\ref{sec:eqset}, one obtains an upper bound of $79^\circ$ on $\gamma$.

If $r_1$ is determined to be smaller, the vertex position will drop lower and
both of the two intercept points of the $A_{CP}(K^* \pi) = 0$ curve will move
toward unity, resulting in a stronger bound on $\gamma$.  For example, $\gamma
\lsim 78^\circ$ if one takes the lower limit $r_1 = 0.22$ in Eq.~(\ref{eq:r1}).

\section{The $\rho K$ Modes}
\label{sec:rhoK}

A similar analysis can be done for the $\rho^- K^+$ and $\rho^+ K^0$ modes.  As
mentioned before, the branching ratio of the $\rho^+ K^0$ mode is yet to be
measured.  One can use Lipkin's assumption instead to evaluate the experimental
value of the ratio $R(\rho^- K^+)$, although this is completely unnecessary if
the $\rho^+ K^0$ decay is seen.  In this case, we have the amplitude
\be
{\cal A}(\rho^- K^+)
= P'_P \left[ 1 + r_2 e^{i (\delta_V + \gamma)} \right] ~,
\ee
where $\delta_V$ is the strong phase of $T'_V$ relative to $P'_V$, and 
\bea
\label{eq:RrhoK}
R(\rho^- K^+)
&=& 1 + 2 r_2 \cos\delta_V \cos\gamma + r_2^2 ~, \\
\label{eq:RACPrhoK}
R(\rho^- K^+) A_{CP}(\rho^- K^+)
&=& 2 r_2 \sin\delta_V \sin\gamma ~.
\eea
One thus obtains the same type of equation as Eq.~(\ref{eq:contour}) for the
$\rho K$ decays.

\begin{figure}[t]
\centerline{\includegraphics[width=8cm]{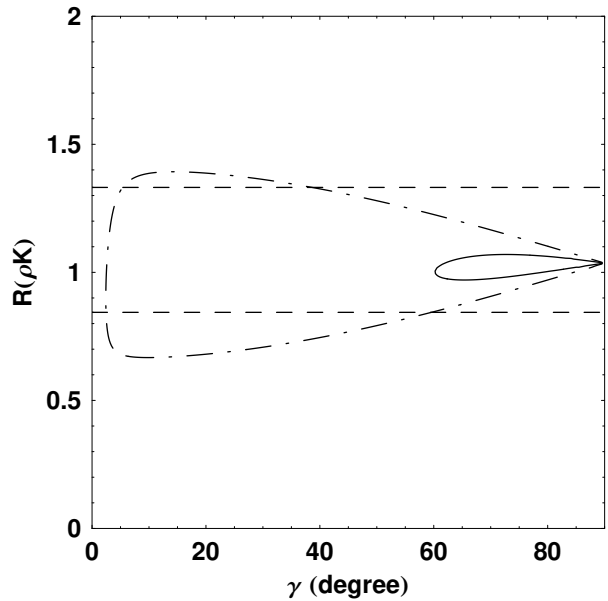}}
\caption[]{Relation between measured $R(\rho K)$ and the weak phase $\gamma$
  for $r_2 = 0.19$.  The solid curve corresponds to $|A_{CP}(\rho K)| = 0.32$,
  and the dash-dotted curve corresponds to $|A_{CP}(\rho K)| = 0.02$.  The
  dashed lines represent the $1\sigma$ range of the observed $R(\rho K)$.
\label{fig:RrhoK}}
\end{figure}

We draw in Fig.~\ref{fig:RrhoK} the curves corresponding to $r_2 = 0.19$.  As
seen in the drawing, the intersection vertex of the asymmetry curves at $\gamma
= 90^\circ$ falls within the $1 \sigma$ limits of $R(\rho K)$, which is
measured to be around unity.  No upper bound on $\gamma$ can be obtained in
such cases.  Since the $CP$ asymmetry is nonzero at $1\sigma$ level, a lower
bound $\gamma \gsim 5^\circ$ is given by the $|A_{CP}(\rho K)| = 0.02$ curve.
This is complementary to the information one learns from the $K^* \pi$ modes in
the previous section, where no lower bound can be obtained due to the observed
$CP$ asymmetry being consistent with zero.  Varying $r_2$ does not change this
lower bound much, and the most conservative bound on $\gamma$ is determined
with the largest possible $r_2$.  Moreover, decreasing $r_2$ shrinks the
$|A_{CP}(\rho K)| = 0.31$ curve toward the vertex at $\gamma = 90^\circ$.

A better determination of $A_{CP}(\rho^- K^+)$ will be able to provide stronger
bounds.  For example, if the upper limit of $A_{CP}(\rho^- K^+)$ is lowered or
$r_2$ turns out to be larger, a middle range of $\gamma$ will be excluded.  On
the other hand, a stronger lower bound on $\gamma$ can be deduced from a larger
lower limit on $A_{CP}(\rho^- K^+)$ or a smaller $r_2$.

\section{The $K^* \pi$ and $\rho K$ Modes Combined}
\label{sec:eqset}

Instead of treating $r_1$ and $r_2$ independently, one may relate one to the
other by employing the relation $|P'_V| = |P'_P|$:
\begin{eqnarray}
\label{eq:rVP}
  r_{VP} \equiv \frac{r_2}{r_1} = \frac{|T'_V|}{|T'_P|}
  = \frac{f_K A_0^{B \rho}(m_K^2)}{f_{K^*} F_1^{B \pi}(m_{K^*}^2)}
  \simeq 0.6 ~,
\end{eqnarray}
where the same numerical factors are used as before.  This can be compared with
the result of $0.7 \pm 0.1$ obtained from a global fit \cite{Chiang:2003pm}.
Therefore, $r_1$ and $r_2$ are related to each other by a simple numerical
factor.  We will take the former as the independent parameter.

There are then four parameters ($\gamma$, $r_1$, $\delta_P$, and $\delta_V$)
for the four observables in Eqs.~(\ref{eq:RKstarpi}), (\ref{eq:RACPKstarpi}),
(\ref{eq:RrhoK}), and (\ref{eq:RACPrhoK}).  One can readily find a solution
$\gamma = 42^\circ$, $r_1 = 0.23$, $\delta_P = 179^\circ$, and $\delta_V =
81^\circ$ from the central values of the observables, where we have again
imposed the requirement that $\gamma$ has to fall within the favored region
($39^\circ - 80^\circ$) determined by Ref.~\cite{Hocker:2001xe}.  (The other
solutions correspond to the transformations $(\gamma,\delta_{P,V}) \to
(\gamma,\pi \pm \delta_{P,V}$) without changing $r_1$.)

It is seen that the value of $\gamma$ extracted using this method rests on the
lower end of the currently preferred range.  The value of $r_1$ is found to be
consistent with our previous estimate.  Note that we do not assume any
knowledge about the size of $T'_P$ or $T'_V$ here and avoid the somewhat
far-reaching relation between them and the tree amplitude in the $\pi \pi$
decays by employing flavor SU(3) symmetry.  Varying $r_{VP}$ between $0.5$ and
$0.7$ does not change the solutions of $\delta_P$ and $\delta_V$ much, but
$\gamma$ and $r_1$ decrease from $48^\circ$ to $37^\circ$ and from $0.25$ to
$0.22$, respectively.  The strong phase $\delta_P = 179^\circ$ means that
$T'_P$ lies almost in line with $P'_P$.  This can be readily understood as the
result of that the central value of $A_{CP}(K^{*+} \pi^-)$ is zero and that
there is a constructive interference between the two amplitudes.

Unfortunately, the uncertainties on the current data are still too large (in
contrast to the $K \pi$ case) to obtain a restricted $1 \sigma$ range for the
weak phase $\gamma$.  If the data precision can be improved by, for example, a
factor of two, a $1 \sigma$ range of $24^\circ - 50^\circ$ can be obtained
assuming the same central values as the present ones.

It should be emphasized that the assumption $|P'_P| = |P'_V|$ can be relaxed
once ${\cal B}(\rho^+ K^0)$ is measured.  As seen in Eq.~(\ref{eq:rVP}), what
is essencial is the factorization of the tree amplitudes.  In that case,
$r_{VP}$ should be further scaled by the factor $|P'_P| / |P'_V|$.  Note that
the information on the relative strong phase between $P'_P$ and $P'_V$ is not
crucial here.

\section{Other Modes}
\label{sec:others}

Let's now turn to the following quantities derived from a larger set of decay
modes:
\begin{eqnarray}
\label{eq:RcKstarpi}
R_c(K^* \pi)
&\equiv& \frac{2\lbar\Gamma(K^{*+} \pi^0)}{\lbar\Gamma(K^{*0} \pi^+)}
= \left| \frac{P'_P + P'_{EW,V} + T'_P + C'_V}{P'_P} \right|^2
< 6.35 ~, \\
\label{eq:RnKstarpi}
R_n(K^* \pi)
&\equiv& \frac{\lbar\Gamma(K^{*+} \pi^-)}{2\lbar\Gamma(K^{*0} \pi^0)}
= \left| \frac{P'_P + T'_P}{P'_P - P'_{EW,V} - C'_V} \right|^2
= 3.77 \pm 2.32 ~, \\
R_c(\rho K)
&\equiv& \frac{2\lbar\Gamma(\rho^0 K^+)}{\lbar\Gamma(\rho^+ K^0)}
= \left| \frac{P'_V + P'_{EW,P} + T'_V + C'_P}{P'_V} \right|^2
> 0.21 ~, \nn \\
\label{eq:RcrhoK}
&\to& \frac{2\lbar\Gamma(\rho^0 K^+)}{\lbar\Gamma(K^{*0} \pi^+)}
= \left| \frac{P'_V + P'_{EW,P} + T'_V + C'_P}{P'_P} \right|^2
= 1.06 \pm 0.23 ~, \\
\label{eq:RnrhoK}
R_n(\rho K)
&\equiv& \frac{\lbar\Gamma(\rho^- K^+)}{2\lbar\Gamma(\rho^0 K^0)}
= \left| \frac{P'_V + T'_V}{P'_V - P'_{EW,P} - C'_P} \right|^2
= 0.97 \pm 0.35 ~.
\end{eqnarray}
Again, ``$\to$'' in Eq.~(\ref{eq:RcrhoK}) indictaes the use of Lipkin's
assumption.  In principle, one can also obtain information on $\gamma$ from the
combination of $R_c(K^* \pi)$ and $A_{CP}(K^{*+} \pi^0)$ and the combination of
$R_c(\rho K)$ and $A_{CP}(\rho^0 K^+)$.  Currently, we are still lacking in
data of the $CP$ asymmetries and two branching ratios.  Moreover, they require
the determination of the corresponding ratios
\begin{eqnarray}
r_{c1} &\equiv& \frac{T'_P + C'_V + P'_{EW,V}}{P'_P} ~, \\
r_{c2} &\equiv& \frac{T'_V + C'_P + P'_{EW,P}}{P'_V} ~.
\end{eqnarray}
This is more involved than the case of $K \pi$ because although $T'+C'+P'_{EW}$
can be deduced from $\pi^+ \pi^0$ using SU(3), the $VP$ counterparts $\rho^+
\pi^0$ and $\rho^0 \pi^+$ contains additional contributions from the penguin
amplitudes $P'_P$ and $P'_V$ that interfere with each other constructively.
Extra assumptions need to be imposed in order to extract the required
information \cite{Chiang:2003pm}.

Recently, it is pointed out that current experimental data indicate some
discrepancy between $R_c(K \pi) = 2\lbar\Gamma(K^+ \pi^0) / \lbar\Gamma(K^0
\pi^+)$ and $R_n(K \pi) = \lbar\Gamma(K^+ \pi^-) / 2\lbar\Gamma(K^0 \pi^0)$
that should be equal to each other at the leading-order expansion.  Such a
discrepancy can be resulted from two possibilities: either the $\pi^0$
detection efficiency in experiments is systematically underestimated, or it
calls for contributions of $\Delta I = 1$ amplitudes beyond the SM
\cite{Gronau:2003kj,Buras:2003dj,Barger:2004hn,Khalil:2004yb,Mishima:2004um,%
  Baek:2004rp}.  The $R_c$-$R_n$ comparison in the cases of $K^* \pi$ and $\rho
K$ decays is somewhat analogous.  However, a direct experimental comparison is
not yet available because of insufficient data in the $K^{*+} \pi^0$ and
$\rho^+ K^0$ modes.  Employing the Lipkin relation again for the $\rho^+ K^0$
mode, the current data show an approximate agreement between
Eq.~(\ref{eq:RcrhoK}) and Eq.~(\ref{eq:RnrhoK}).

If the discrepancy between $R_c$ and $R_n$ in the $K \pi$ system is due to the
underestimated $\pi^0$ detection efficiency, we expect a similar pattern in the
$K^* \pi$ system but not in the $\rho K$ modes.  This seems to be partly
favored by the rough agreement between Eq.~(\ref{eq:RcrhoK}) and
Eq.~(\ref{eq:RnrhoK}), although a mild assumption is used here and the $K^*
\pi$ counterparts still await to be seen.  Suppose the interpretation of new
physics, which is short-distance in nature, is correct for the $K \pi$ modes,
then one should expect to see its effects in the $VP$ (both $K^* \pi$ and $\rho
K$) and $VV$ ($K^* \rho$) modes too.  However, a detailed analysis is required
because the latter modes involve different polarizations.

Finally, we would like to comment on possible branching ratios of the $K^{*+}
\pi^0$ and $\rho^+ K^0$ decays by noting two approximate sum rules:
\begin{eqnarray}
  \label{eq:sumrule1}
  \lbar\Gamma(K^{*0} \pi^+) + \lbar\Gamma(K^{*+} \pi^-)
  &\approx&
  2 \left[ \lbar\Gamma(K^{*+} \pi^0) + \lbar\Gamma(K^{*0} \pi^0) \right]
  ~, \\
  \label{eq:sumrule2}
  \lbar\Gamma(\rho^+ K^0) + \lbar\Gamma(\rho^- K^+)
  &\approx&
  2 \left[ \lbar\Gamma(\rho^0 K^+) + \lbar\Gamma(\rho^0 K^0) \right]
  ~.
\end{eqnarray}
They hold only when the terms $|C'_{V(P)} + P'_{EW,V(P)}|^2 + 2 {\rm
  Re}[T^{\prime *}_{P(V)} (C'_{V(P)} + P'_{EW,V(P)})]$ are negeligible in
comparison with the dominant penguin contributions.  It is noticed from a
global fit to $VP$ data \cite{Chiang:2003pm} that the contributions due to
$C'_V$ and $P'_{EW,V}$ are quite sizeable, thus raising doubt on the sum rule
(\ref{eq:sumrule1}) while those due to $C'_P$ and $P'_{EW,P}$ are less
significant.  Assuming these sum rules, one can deduce from current data that
${\cal B}(K^{*+} \pi^0) = (9.9 \pm 1.6) \times 10^{-6}$ and ${\cal B}(\rho^+
K^0) = (10.7 \pm 4.3) \times 10^{-6}$.  These numbers are consistent with the
current upper bounds.  In particular, ${\cal B}(\rho^+ K^0)$ thus obtained and
the measured ${\cal B}(K^{*0} \pi^+)$ are about the same, which is another
indication of the equality $|P'_P| = |P'_V|$.  At any rate, a precise
determination of the rates of $K^{*+} \pi^0$ and $\rho^+ K^0$ decays will be
very helpful in checking the $R_c$-$R_n$ relations and the above sum rules.

\section{Conclusions}
\label{sec:conclusions}

We have shown that the branching ratios and $CP$ asymmetries of the $K^* \pi$
and $\rho K$ modes can help us constraining the weak phase $\gamma$.  In
particular, we emphasize the uses of the $K^{*0} \pi^+$, $K^{*+} \pi^-$,
$\rho^+ K^0$, and $\rho^- K^+$ decays.  With the choice of $r_1 = 0.32$ and
$r_2 = 0.19$, we show that the existing data give us the bounds $\gamma \lsim
86^\circ$ from the $K^* \pi$ modes and $\gamma \gsim 5^\circ$ from the $\rho K$
modes.  Although currently very loose, these bounds are expected to improve
with higher data precision in the coming years.

By relating the two types of tree amplitudes, $T'_P$ and $T'_V$, in the $VP$
decays, one can reduce the number of parameters in the problem and solve for
$\gamma$.  This method is free from SU(3) breaking uncertainties.  It is found
that the solution of $\gamma$ thus obtained is consistent with those obtained
from other observables.  However, a better constraint on $\gamma$ from these
modes is not possible until the data precision is further improved.  The
solution also tells us the tree-to-penguin ratio, which we find to agree with
that estimated using flavor symmetry.

We also point out that it is interesting to check whether $R_c = R_n$ for both
$K^* \pi$ and $\rho K$ decays.  As shown above, the equality roughly holds for
the $\rho K$ modes when Lipkin's assumption is used.  If the equality turns out
to hold in the case of $\rho K$ but to be violated for $K^* \pi$ with a
similarly puzzling difference as the $K \pi$ decays, then it is likely that the
underestimated $\pi^0$ detection efficiency explanation is favored.  On the
other hand, if the data tell us that both equalities are violated in these $VP$
modes, then the new physics explanation is more plausible.  However, one then
has to work out the new physics contributions to different polarization
components.  We therefore stress the importance of measuring the yet unseen
modes in order for us to validate the use of the relation $P'_V = -P'_P$ and to
see any further hints on new physics.

\vspace{0.2cm}

{\bf Acknowledgments:} The author wishes to thank Paoti Chang for useful
information on experimental issues and Jonathan Rosner for comments on the
manuscript.  He is also grateful for the kind hospitality of National Center
for Theoretical Sciences in Taiwan during his visit when part of the work was
done.  This work was supported in part by the National Science Council of
Taiwan, R.~O.~C.\ under Grant No.\ NSC 93-2119-M-008-028-.


\end{document}